\documentclass[11pt]{article}
\usepackage[dvips]{graphicx}
\usepackage{color}
\newcommand {\be}{\begin{eqnarray}}
\newcommand {\ee}{\end{eqnarray}}

\newcommand {\nd}{\noindent}
\begin{document}
\begin{center}
{\large \bf Effect of geometrical size of the particles in a hot and dense hadron gas}
\vskip 0.2in
{\bf M. Mishra\footnote{madhukar.12@gmail.com} and C. P. Singh\footnote{cpsingh\_bhu@yahoo.co.in}}\\
{\it Department of physics, Banaras Hindu University,
Varanasi - 221005, India}
\vskip 0.2in
{\bf ABSTRACT}
\end{center}
\vskip 0.2in

    Incorporation of the finite size of baryons into the equation of state (EOS) of a hot and dense hadron gas (HG) in a thermodynamically consistent manner has been a much studied problem. We first review its current status. Various models have been proposed in order to account for the repulsive force generated by the hard-core geometrical size of the baryons resulting in an excluded volume effect in the EOS. We examine the criterion of the thermodynamical consistency of these models and summarize their shortcomings. In order to remove the shortcomings, we propose a new model which incorporates the excluded volume effect in a thermodynamically consistent manner. We find that the new model works even for the cases of extremely large temperatures and densities where most of other approaches fail. Furthermore, the new expressions for thermodynamical variables resemble in form with those obtained from thermodynamically inconsistent models and thus a useful correction factor has been suggested here which converts inconsistent expressions into thermodynamically consistent ones. Finally we compare the predictions of new model with those obtained from various old models.
\vskip 0.4in

{\nd \bf PACS numbers}: 25.75.-q, 21.65.+f, 24.10.Pa

{\nd \bf Keywords}: Hot and dense hadron gas, Equation of state, Thermodynamical consistency, Geometrical hard-core volume of baryons, Causality, Quark-Gluon Plasma,
Phase transition.

\newpage

\section{INTRODUCTION}
   Quantum chromodynamics (QCD) predicts a phase transition at a large temperature and/or large baryon density from a normal colour confined phase of hadronic matter to a deconfined phase of Quark Gluon Plasma (QGP). Many aspects of this phase transition are still poorly understood [1-6] and are under intense investigations. Therefore, it is worthwhile to precisely determine the properties of hot and dense hadronic matter in order to devise some unique signals of QGP [7] formation and to determine the QGP properties [2,3]. It requires to intensify the search for a proper equation of state (EOS), which can suitably describe the properties of a hot and dense hadron gas (HG). The determination of nuclear matter equation of state (EOS) at very large temperature and density still remains as one of the most significant goals in both theoretical and experimental investigations which are being pursued in heavy-ion physics and cosmology.

    We want to explore the properties of hadronic matter in unusual environments, in particular at large temperatures and/or high baryon densities. There are many reasons for such an investigation. First one might hope to find in nature or to produce in laboratory such extreme conditions and hence one can test the theory in this new domain. There are three places where one might look for hadonic matter at high temperatures and/or large baryon densities. The standard cosmological models allow one to extrapolate back to about 10 $\mu\,s$ after the big bang when the universe as a whole was at a very high temperature. Similarly the interior of the neutron stars is expected to contain a matter with significantly higher baryon density matter than the normal nuclear density. One can also produce such a matter at extremely high temperatures and/or density in a laboratory by colliding two heavy-ions at very high energy density. In all the above cases, understanding of the physics of this unusual matter requires knowledge of the EOS and the properties of the hadronic matter. In a simple treatment of the HG, all the baryons are treated as non-interacting, point-like objects. Such an EOS of the HG has an undesirable feature that at very high baryon densities or chemical potentials, the hadronic phase reappears as a stable configuration in the Gibbs construction of the phase equilibrium between the HG and QGP phases. For small $\mu_B$, the pressure in the QGP phase is smaller than in the HG phase because of the negative vacuum pressure $-B$ present in the QGP phase and thus the hadronic phase is stable. However, in general the QGP pressure increases with $\mu_B$ and/or $T$ faster than the hadronic pressure. This is due to the fact that QGP phase has far more degrees of freedom than the HG. Thus at a fixed $T$, we get a critical chemical potential at which $P_{HG}=P_{QGP}$ and the transition to the QGP phase takes place. However, at still higher $\mu_B$, the hadronic phase possesses still larger degrees of freedom due to an exponential growth of hadrons and resonances resulting in a higher pressure for HG than QGP. This signifies a reversal of phase transition from QGP to HG. This is an anomalous situation [8-10] since we know the stable phase at any given $(T,\,\mu_B)$ is the phase which has a larger pressure. However, one expects that once the system goes over to the QGP phase, it should remain in that phase even at extremely large $\mu_B$ due to the property of asymptotic freedom of QCD. Moreover, it is expected that the hadronic interactions become significant when hadrons are closely packed in a hot and dense hadron gas. This anomalous behaviour arises because HG has been considered as an ideal gas of non-interacting, point-like hadrons. As a result of this assumption, the thermal production of an arbitrarily large number of hadrons in a given volume at large $\mu_B$ or $T$ is possible and eventually leads to infinitely large energy densities and pressures. A simple remedy of the above problem is provided by the inclusion of a finite, proper volume for each baryon, which leads to a hard-core repulsion among them at very high density and/or temperature. Many phenomenological models have been invented for this purpose [11-24] in the most recent past. It puts a maximum limit or bound for the number of hadrons in fireball so that its volume is completely filled with particles. This repulsive force has been incorporated in the literature by giving a geometrical hard-core volume to each baryon and it is more commonly known as excluded volume effect [11-21].

     Mean field theoretic models [25,26] and their phenomenological generalizations [27-30] constitute another important approach for the construction of an EOS for the HG phase. In these models, one starts from local renormalizable Lagrangian densities with baryonic and mesonic degrees of freedom. These models rigorously satisfy causality (i.e., velocity of sound $v_s$ in the medium is smaller than velocity of light). They also reproduce the ground state properties of the nuclear matter in the low-density limit. The short-range repulsive interaction in these models arises due to $\omega$-exchange between a pair of baryons. It leads to the Yukawa potential $V(r)=(G^2/4\,\pi\,r)\,exp(-m_{\omega}\,r)$ , which further gives mean potential energy of HG as $U_B=G^2\,n_B/m_{\omega}$, which means that $U_B$ is proportional to the net baryon density $n_B$. Thus $U_B$ vanishes in the $n_B\rightarrow 0$ limit. In the baryon less limit, hadrons (mesons) can still approach point-like behaviour due to the vanishing of the repulsive interaction between them. It means that in principle one can excite a large number of hadronic resonances at large $T$. This will again make the pressure in the HG phase larger than the pressure in the QGP phase and the hadronic phase would again become stable at sufficiently high $T$ and we will not get a reasonable phase transition according to the Gibbs construction. In some recent approaches this problem has been cured by considering another temperature dependent mean-field $U_{VDW}(n,T)$, where $n$ is the sum of particle and anti-particle number densities. Here $U_{VDW}(n,T)$ represents Vander-Waals hard-core repulsive interaction between two particles and depends on the total number density $n$ and is non-zero even when net baryon density $n_B$ is zero in the large temperature limit. However, in the high-density limit, the presence of a large number of hyperons and their unknown couplings to the mesonic fields generates a large amount of uncertainty in the EOS of HG in the mean-field model. Moreover, we find that the EOS formulated in the hot, dense scenario usually suffers from a crucial assumption regarding how many particles and resonances one should incorporate into the EOS for a realistic description of HG. The mean-field models can usually handle very few resonances only in the description of HG and hence are not very reliable.

    The purpose of this paper is to review the status of excluded volume models used in the literature. We will then point out the problem of thermodynamic inconsistency in the models and then describe the shortcomings present in the thermodynamically consistent models. Here we propose a new thermodynamically consistent excluded volume model and examine its predictions and compare with others.

\section{EXCLUDED VOLUME MODELS}

    The net excluded pressure, number density and the energy density of a multi-component
HG are given in a simple model by Cleymans and Suhonen [11] as:
\begin{equation}
P^{ex}=\frac{\sum_i\,P_i^0}{1+\sum_i\,n_i^0\,V_i^0},
\end{equation}
\begin{equation}
n^{ex}=\frac{\sum_i\,n_i^0}{1+\sum_i\,n_i^0\,V_i^0},
\end{equation}                        
\begin{equation}
\epsilon^{ex}=\frac{\sum_i\,\epsilon_i^0}{1+\sum_i\,n_i^0\,V_i^0},
\end{equation}                               
where $P_i^0,\;\epsilon_i^0,\;n_i^0\;$ and $V_i^0$ are pressure, energy density, net baryon density and eigen volume of $i^{th}$ baryon, respectively. Here $\Sigma_i\,n_i^0\,V_i^0$ is the fraction of occupied volume. Kuono and Takagi [14] modified these expressions by considering the existence of a repulsive interaction either between a pair of baryons or between a pair of anti-baryons. Therefore, the expressions (1), (2) and (3) take the form:
\begin{equation}
n^{ex}=\frac{\sum_i\,n_i^0}{1+\sum_i\,n_i^0\,V_i^0}-\frac{\sum_i\,n_{\bar i}^0}
{1+\sum\,n_{\bar i}^0\,V_i^0},
\end{equation}
\begin{equation}
P^{ex}=\frac{\sum_i\,P_i^0}{1+\sum_i\,n_i^0\,V_i^0}-\frac{\sum_i\,P_{\bar i}^0}
{1+\Sigma_i\,n_{\bar i}^0\,V_i^0}+P_M^0,
\end{equation}
\begin{equation}
\epsilon^{ex}=\frac{\sum_i\,\epsilon_i^0}{1+\sum_i\,n_i^0\,V_i^0}+\frac{\sum_i\,\epsilon_{\bar i}^0}{1+\sum_i\,n_{\bar i}^0\,V_i^0}+\epsilon_M^0,
\end{equation}                                                                             
where $n_i^0$ and $n_{\bar i}^0$ are the number density of the point-like baryons and anti-baryons respectively, $\epsilon_i^0(\epsilon_{\bar i}^0)$ and $P_i^0(P_{\bar i}^0)$  are the corresponding energy density and pressure. Here, $P_M^0,\;\epsilon_M^0$ are the pressure and energy density of point-like mesons.

    In the Hagedorn model [15], the excluded volume correction is proportional to the point-like energy density $\epsilon^0$. The grand canonical partition function satisfies   
$$
\begin{array}{lcl}
ln\,Z(T,V,\lambda)=ln\,Z^0(T,\Delta,\lambda),
\end{array}
$$    
where it has been assumed that the density of states of the finite-size particles in total volume $V$ is the same as that of point-like particles in the available volume $\Delta$ where $\Delta=V-\Sigma_i\,V_i^0$. The sum of eigen volumes $\Sigma_i\,V_i^0$ is given by the ratio of the invariant cluster mass to the total energy density and $\lambda$ is the fugacity i.e., $\lambda=exp(\mu/T)$. Hence $\Sigma_i\,V_i^0=E/4\,B=V\,\epsilon/4\,B$ and $\epsilon=\Delta\,\epsilon^0/V$. Correcting for the factor $\Delta$, one finally gets:
\begin{equation}
\epsilon_i^{ex}=\frac{\epsilon_i^0}{1+\epsilon^0/4\,B}.
\end{equation}                                                                     
The Limiting case of $\mu\rightarrow \infty$ yields $\epsilon^0/4\,B\gg 1$ and $\epsilon=\Sigma_i\,\epsilon_i^{ex}=4\,B$ which is obviously the upper limit for $\epsilon$as it gives the energy density existing inside a nucleon and usually regarded as the latent heat density required for the phase transition. Here $B$ is the bag constant. The number density and pressure can be written similarly in this model as:
\begin{equation}
n_i^{ex}=\frac{n_i^0}{1+\epsilon^0/4\,B},
\end{equation}
\begin{equation}
P_i^{ex}=\frac{P_i^0}{1+\epsilon^0/4\,B}.
\end{equation}                                                                      However, the main drawback existing in all the above models is the lack of the consistency with the basic thermodynamic relations. This implies that $n_B\neq\partial\Omega/\partial\mu_B$ i.e., the net baryon density in principle cannot be derived from a thermodynamic potential $\Omega$. The question of thermodynamic consistency was first examined in detail by Rischke et al. [16]. The grand canonical partition function $Z_G$ for point-like baryons was written in terms of canonical partition function $Z_C$ as:
\begin{equation}
Z_G^0(T,\mu,V)=\sum_{N=0}^{\infty}\,exp(\mu\,N/T)\,Z_C(T,N,V).
\end{equation}
They further modified the canonical partition function $Z_C$ by introducing a step-function in the volume so as to incorporate excluded volume correction into the formalism. Therefore, the grand canonical partition function (10) finally takes the form:
\begin{equation}
Z_G^{ex}(T,\mu,V-V^0\,N)=\sum\,exp(\mu\,N/T)\,Z_C(T,N,V-V^0\,N)\,\theta(V-V^0\,N).
\end{equation}                             
Using the Laplace transform, one gets the isobaric partition function as:
\begin{equation}
Z_P=\int_0^{\infty}dV\,exp(-\xi\,V )Z_G^{ex}(T,\mu,V-V^0\,N).
\end{equation}                                                                          
Or one gets after rearrangement of the terms:
\begin{equation}
Z_P=\int_0^{\infty}dx\,exp\left\{-x\left[\xi-\frac{ln\,Z_G^0(T,\tilde\mu)}{x}\right]\right\},
\end{equation}                                                                           
where $x=V-V^0\,N$ and $\tilde\mu=\mu-T\,V^0\,\xi$. Finally, we get a transcendental equation as follows:
\begin{equation}
P^{ex}(T,\mu)=P^0(T,\tilde\mu),
\end{equation}                                                                     
where
\begin{equation}
\tilde{\mu}=\mu-V^0\,P^{ex}(T,\mu).
\end{equation}                                                                   
The expressions for number density, entropy density, and energy density in this formalism can be obtained as:
\begin{equation}
n^{ex}(T,\mu)=\frac{\partial P^0(T,\tilde\mu)}{\partial\tilde\mu}\,\frac{\partial\tilde\mu}{\partial\mu}=\frac{n^0(T,\tilde\mu)}{1+V^0\,n^0(T,\tilde\mu)},
\end{equation}
\begin{equation}
s_1^{ex}(T,\mu)=\frac{\partial P^0(T,\tilde\mu)}{\partial T}=\frac{s_1^0(T,\tilde\mu)}{1+V^0\,n^0(T,\tilde\mu)},
\end{equation}
\begin{equation}
\epsilon^{ex}(T,\mu)=\frac{\epsilon^0(T,\tilde\mu)}{1+V^0\,n^0(T,\tilde\mu)}.
\end{equation}                                                                     
These equations resemble with Eqs. (2), and (3) as given in Cleymans-Suhonen model. Here $\mu$ has been replaced by $\tilde\mu$. The above model can be extended for a hadron gas involving several baryonic species. Considering an ideal hadron gas consisting of several baryonic species $i=1,\cdots,h$, the thermodynamic functions are additive and equal to the sum of their partial values of different particle species as. So we find:
\begin{equation}
P^{ex}(T,\mu_1,\cdots,\mu_h)=\sum_{i=1}^h\,P_i^0(T,\tilde\mu_i),
\end{equation}                                                                            where
\begin{equation}
\tilde{\mu_i}=\mu_i-V_i^0\,P^{ex}(T,\mu_i),
\end{equation}         
with $i=1,\cdots,h$. Particle number density for the $i^{th}$ species can be calculated from Eq. (20) and Eq. (21)
\begin{equation}
n_i^{ex}(T,\mu_i)=\frac{n_i^0(T,\tilde\mu_i)}{1+\sum_{j=1}^h\,V_j^0\,n_j^0(T,\tilde\mu_j)}.
\end{equation}                                                                            
Unfortunately, the above model involves cumbersome, transcendental expressions which are usually not easy to calculate. In solving Eq. (14) and (15), however, one can use a trick. First we choose a value for $\tilde\mu$ and evaluate $P^{ex}$ from Eq. (14) and then one can find out the value of $\mu$ from (15) provided one knows the eigen volume $V^0$.

    Singh et al. [17] have also proposed a thermodynamically consistent model for the inclusion of excluded volume correction. Using Boltzmann approximation, one can write the partition function as follows:
\begin{equation}
ln\,Z_i^{ex}=\frac{g_i\,\lambda_i}{6\,\pi^2\,T}\,\int_{V_i^0}^{V-\sum_j\,N_j\,V_j^0}
dV\,\int_0^{\infty}\,\frac{k^4\,dk}{\sqrt{k^2+m_i^2}}\,exp(-\sqrt{k^2+m_i^2}/T),
\end{equation}                                                 
where $\lambda_i$ is the fugacity of $i^{th}$ component of baryonic species, $k$ is the magnitude of the momentum of hadrons and $N_j$ is the total number of $j^{th}$ type of baryons. We can rewrite it as:
\begin{equation}
ln\,Z_i^{ex}=V(1-\sum_j\,n_j^{ex}\,V_j^0)\,I_i\,\lambda_i,
\end{equation}                                                                             where integral $I_i$ is
\begin{equation}
I_i=\frac{g_i}{2\,\pi^2}\left(\frac{m_i}{T}\right)^2\,T^3\,K_2\,(m_i/T).
\end{equation}
Therefore, the partition function has been directly corrected for the excluded volume effect.
The number density of $i^{th}$ baryonic species can be obtained from $Z_i^{ex}$ as:
$$
\begin{array}{lcl}
n_i^{ex}=\frac{\lambda_i}{V}\left(\frac{\partial\,ln\,Z_i^{ex}}{\partial\, \lambda_i}\right)_{T,V}.
\end{array}
$$
We get from Eq. (23) as:
\begin{equation}
n_i^{ex}=(1-R)\,\lambda_i\,I_i-I_i\,\lambda_i^2\,\frac{\partial R}{\partial \lambda_i}.
\end{equation}                                                          
Here we define $R=\sum_i\,n_i^{ex}\,V_i^0$. It is obvious that the thermodynamically inconsistent expressions (1) and (2) can be obtained from Eq. (25) if we put the factor $\partial R/\partial \lambda_i=0$ and consider only one type of baryons in the system. Thus the presence of $\partial R/\partial \lambda_i$ in Eq. (25) corrects for the inconsistency. For single component HG, one gets the solution of Eq. (25) as:
\begin{equation}
n^{ex}=\frac{1}{V}\,\frac{\int_0^{\lambda}exp[-1/I\,V^0\,\lambda]}{\lambda\,exp[-1/I\,V^0\,\lambda]}.
\end{equation}                                             
For a multi-component hadron gas, Eq. (25) can be put in the form
\begin{equation}
R=(1-R)\sum_i I_i\,V_i^0\,\lambda_i-\sum_i I_i\,V_i^0\,\lambda_i^2\,\frac{\partial R}{\partial\lambda_i}.
\end{equation}                                                                             
Using the method of parametric space, one can define [18]:
\begin{equation}
\lambda_i=\frac{1}{(a_i\,+I_i\,V_i^0\,t)}.
\end{equation}                
We finally get the solution of Eq. (28) as follows:
\begin{equation}
R=1-\frac{\int_t^{\infty}[exp(-t^{'})/G(t^{'})]dt^{'}}{exp(-t)/G(t)},
\end{equation}                                                                             where $t$ is a parameter such that
$$
\begin{array}{lcl}
d\lambda_i(t)/dt=-I_i\,\lambda_i^2\,V_i^0,
\end{array}
$$                                               
\begin{equation}
G(t)=t\,\sum_{i=2}^h(a_i+I_i\,V_i^0\,t).
\end{equation}                                                                             If $\lambda_i$' s and $t$ are known, one can determine $a_i$' s. The quantity $t$ is fixed by setting $a_1=0$ and one obtains $t=1/I_1\,V_1\,\lambda_1$, here the subscript 1 refers to the nucleonic degree of freedom and $h$ is the total number of baryonic species. Hence by using $R$ and $\partial R/\partial\lambda_i$ one can calculate $n_i$. It is obvious that the above solution is not unique. Since it contains some parameters $a_i$, the value of one of them has been fixed to zero  arbitrarily. Alternatively, one can assume [17]:
\begin{equation}
\frac{\partial R}{\partial\lambda_i}=\frac{\partial \sum_j n_j^{ex}\,V_j^0}{\partial\lambda_i}=\left(\frac{\partial n_i^{ex}}{\partial\lambda_i}\right)\,V_i^0.
\end{equation}                                                                             Here it has been assumed that the number density of $i^{th}$ baryon will only depend on the fugacity of same baryon. Then the Eq. (27) reduces to the following simple form
\begin{equation}
\frac{\partial n_i^{ex}}{\partial\lambda_i}+n_i^{ex}\left(\frac{1}{I_i\,V_i^0\,\lambda_i^2}+\frac{1}{\lambda_i}\right)=\frac{1}{\lambda_i\,V_i^0}\left(1-\sum_{i\neq j}\,n_j^{ex}\,V_j^0\right).
\end{equation}                                                           
The solution of Eq. (32) can then be obtained in a straightforward manner [17]:
\begin{equation}
n_i^{ex}=\frac{Q_i(1-\sum_{j\neq i}\,n_j^{ex}V_j^0)}{\lambda_i\,V_i^0}\,
exp\,(1/I_i\,V_i^0\,\lambda_i),
\end{equation}
where
$$
\begin{array}{lcl}
Q_i=\int_0^{\lambda_i}exp(-1/I_i\,V_i^0\,\lambda_i)\,d\lambda_i.
\end{array}
$$
In this model R can be obtained by using the relation
\begin{equation}
R=\sum_j n_j^{ex}\,V_j^0=\frac{X}{1+X},
\end{equation}                                                                       
where
\begin{equation}
X=\frac{\sum_i n_i^{ex}\,V_i^0}{1-\sum_i n_i^{ex}\,V_i^0}.
\end{equation}                                                                             Here $X$ is the ratio of the occupied to the available volume. Finally, the number density of $i^{th}$ baryonic species can be written as:
\begin{equation}
n_i^{ex}=\frac{(1-R)}{V_i^0}\,\frac{Q_i}{\lambda_i\,exp(-1/I_i\,V_i^0\,\lambda_i)-Q_i}.
\end{equation}
It is obvious from Eq. (36) that we have obtained an easy and simple solution. There is no parameter in this theory and thus it can be regarded as a unique solution. However, this still depends crucially on the assumption that the number density of $i^{th}$ species is a function of the $\lambda_i$ alone and it is independent of the fugacities of other kinds of baryons. As the interactions between different species become significant in hot and dense HG, this assumption is no longer valid. Moreover, one serious problem crops up, as we cannot do calculation in this model for $T<185$ MeV (and $\mu_B>450$ MeV). This particular limiting value of temperature and baryon chemical potential depends significantly on the masses and the degeneracy factors of the baryonic resonances considered in the calculation.

   In order to remove the above discrepancies, we propose here a new model by rewriting Eq. (27) as:
\begin{equation}
R=(1-R)\sum_i n_i^0\,V_i^0-\sum_i n_i^0\,V_i^0\,\lambda_i\frac{\partial R}{\partial \lambda_i}.
\end{equation}   
Taking $R^0=\sum_i A_i$, where $A_i=I_i\,\lambda_i\,V_i^0$ which means $R^0=\sum_i n_i^0\,V_i^0$ and putting $\partial R/\partial\lambda_i=0$, we get
\begin{equation}
R=\hat R=\frac{R^0}{1+R^0}. 
\end{equation}                                                                     
Then Eq. (37) can be cast in a simplified form:
\begin{equation}
R=\hat R+\Omega R.
\end{equation}                                                                         where the operator $\Omega$ has the following form
\begin{equation}
\Omega=-\frac{1}{1+R^0}\sum_i I_i\,\lambda_i^2\,V_i^0\,\frac{\partial}{\partial\lambda_i}.
\end{equation}                                                                            By using Neumann iteration method, Eq. (37) can be written in a series form as:
\begin{equation}
R=\hat R+\Omega\hat R+\Omega^2\hat R+\Omega^3\hat R \cdots\cdots
\end{equation}                                                                            By using the term upto second order (i.e., $\Omega^2\hat R$), we find
\begin{equation}
R=\frac{\sum_i\,A_i}{1+\sum\,A_i}-\frac{\sum_i\,A_i^2}{(1+\sum_i\,A_i)^3}+
2\frac{\sum_i\,A_i^3}{(1+\sum_i\,A_i)^4}-3\frac{\sum_i\,A_i\,\lambda_i\sum_i\,
A_i^2\,I_i\,V_i}{(1+\sum_i\,A_i)^5}.
\end{equation}                                  
In Eq. (42), we find that the first term yields the value of $R$ in inconsistent model. So other term in Eq. (42) are the correction terms required by thermodynamic consistency.
Finally by calculating the values of $R$ and $\partial R/\partial\lambda_i$, one can calculate the values of particle number density by using Eq. (25).  Similarly the baryonic pressure can be given as:
\begin{equation}
P^{ex}=(1-R)\sum_i P_i^0.
\end{equation}                                                                            We have calculated the energy densities of all the baryons numerically by using the expression
\begin{equation}
\epsilon^{ex}=\frac{T^2}{V}\frac{\partial\,ln\,Z_i^{ex}}{\partial\,T}+\mu_i\,n_i^{ex}.
\end{equation}                                                                             Similarly entropy density of the hadrons can also be calculated from the expression
\begin{equation}
s=\frac{\epsilon^{ex}+P^{ex}-\mu_B\,n_B-\mu_s\,n_s}{T}.
\end{equation}
        
   Obviously this approach looks more simple and attractive in comparison to other excluded volume approaches which are thermodynamically consistent. Moreover, this approach has an added advantage as it can be used for extremely low as well as extremely large values of temperature $T$ and baryon chemical potential $\mu_B$ where all the other approaches either fail to give a satisfactory result or become cumbersome to calculate.
   
   We have considered all baryons and mesons and their resonances having masses up to $2$  GeV/$c^2$ in our calculation. In order to conserve strangeness quantum number, we have used the criteria of equating the net strangeness number density equal to zero as:
\begin{equation}
\sum_i\,S_i(n_i^s-n_{\bar i}^s)=0,
\end{equation}   
where $S_i$, is the strangeness quantum number of $i^{th}$ hadron, $n_i^s$ and $n_{\bar i}^s$ are the strangeness density of $i^{th}$ hadron and $i^{th}$ anti-hadron, respectively. In all the above calculations we have considered mesons behaving as point-like particles. Furthermore, we have taken an equal eigen volume for each baryonic component as $V^0=4\,\pi\,r^3/3$ and a hard-core radius $r=0.8$ fm.

\begin{figure}[tbp]
\begin{center}
\mbox{\includegraphics[scale=0.8]{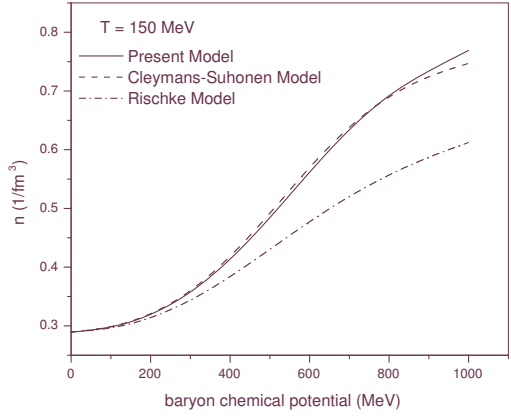}}
\caption{Total number density $n$ versus $\mu_B$ at constant temperature $T=150$ MeV calculated by our model, inconsistent model and Rischke model.}
\end{center}
\end{figure}
\begin{figure}[tbp]
\begin{center}
\mbox{\includegraphics[scale=0.8]{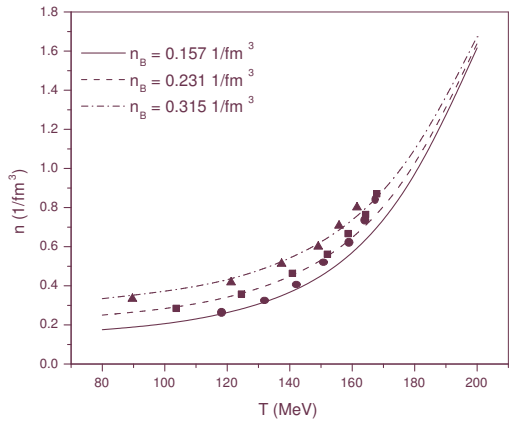}}
\caption{Total number density $n$ versus temperature $T$ at constant net baryon
Density $n_B$. We have also shown the total number density as calculated by Sasaki using an event generator by solid points for the comparison with our model results.}
\end{center}
\end{figure}
\begin{figure}[tbp]
\begin{center}
\mbox{\includegraphics[scale=0.8]{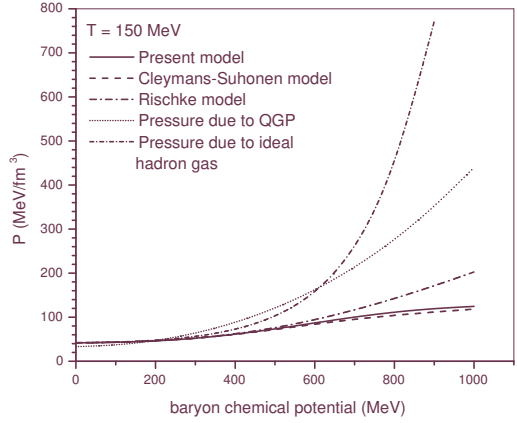}}
\caption{Variation of total hadronic pressure versus $\mu_B$ is shown at constant temperature $T=150$ MeV as calculated in our present model, Cleymans-Suhonen model and Rischke model. We have also plotted pressure for QGP calculated from Eq. (47) (Bag model EOS) with respect to $\mu_B$ at $T=150$ MeV. We have also shown the pressure arising due to ideal hadron gas in the figure.}
\end{center}
\end{figure}
\begin{figure}[tbp]
\begin{center}
\mbox{\includegraphics[scale=0.8]{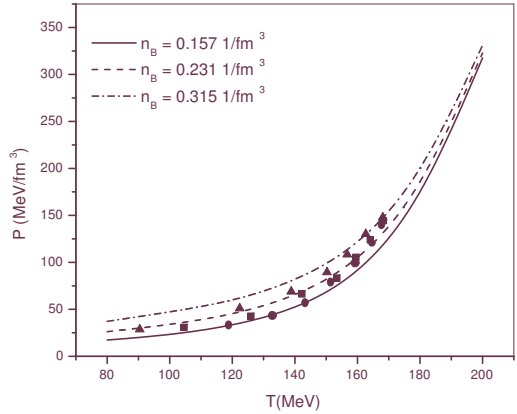}}
\caption{Variation of hadronic pressure $P$ with respect to temperature $T$ at constant baryon density $n_B$. We have also shown hadronic pressure calculated by Sasaki using event generator by solid points for the comparison with our result.}
\end{center}
\end{figure}
\begin{figure}[tbp]
\begin{center}
\mbox{\includegraphics[scale=0.8]{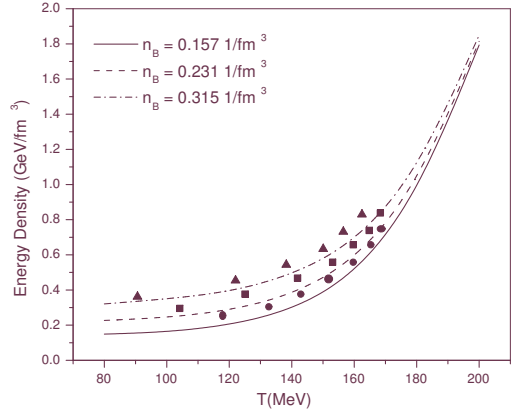}}
\caption{Variation of energy density $\epsilon$ with respect to temperature $T$ at constant baryon density $n_B$. We have also plotted energy density as calculated by Sasaki by solid points for comparison.}
\end{center}
\end{figure}
\begin{figure}[tbp]
\begin{center}
\mbox{\includegraphics[scale=0.8]{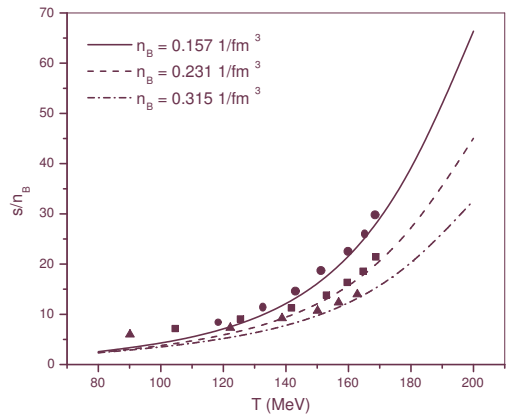}}
\caption{Variation of the $s/n_B$ with respect to temperature $T$ at constant $n_B$ as calculated by our model as well as by Sasaki using event generator.}
\end{center}
\end{figure}
\begin{figure}[tbp]
\begin{center}
\mbox{\includegraphics[scale=0.8]{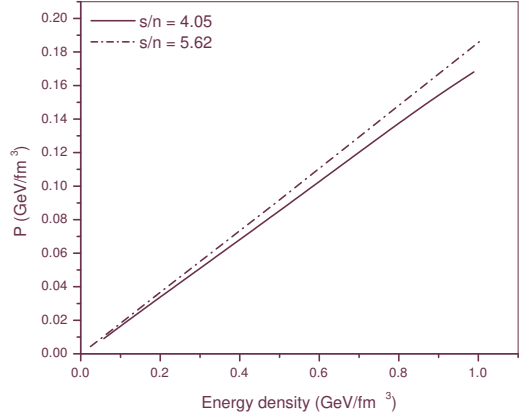}}
\caption{Pressure versus energy density at constant $s/n$ calculated in our model.}
\end{center}
\end{figure}
\begin{figure}[tbp]
\begin{center}
\mbox{\includegraphics[scale=0.8]{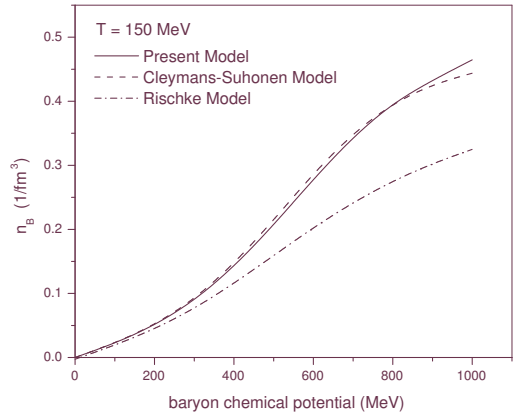}}
\caption{Variation of the baryon density $n_B$ with baryon chemical potential $\mu_B$ at constant temperature $T=150$ MeV.}
\end{center}
\end{figure}
\begin{figure}[tbp]
\begin{center}
\mbox{\includegraphics[scale=0.8]{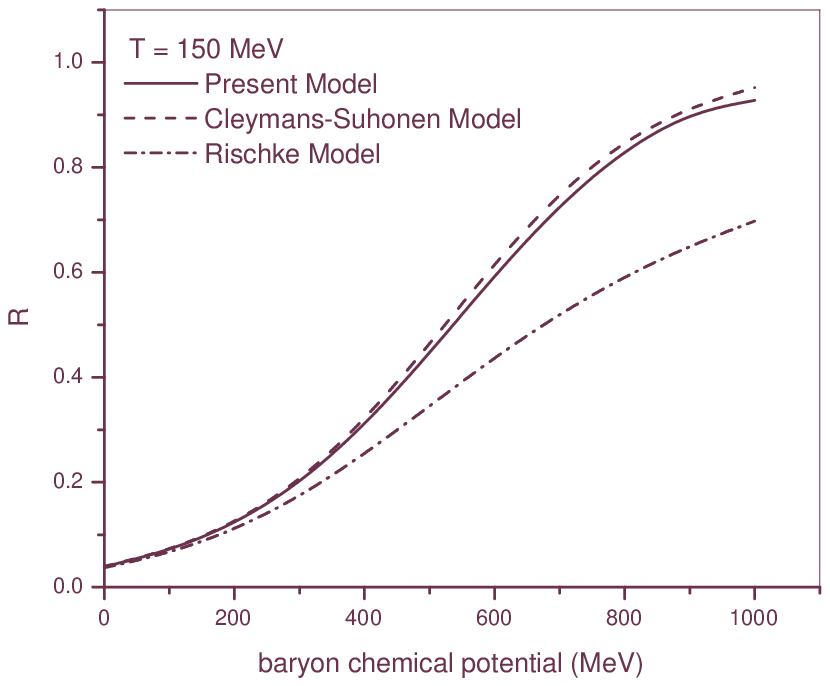}}
\caption{Plot of fraction of occupied volume $R$ versus baryonic chemical potential $\mu_B$ at constant temperature $T=150$ MeV calculated by our model, inconsistent model as well as Rischke model}
\end{center}
\end{figure}

\section{RESULTS AND DISCUSSIONS}

    Figure 1 shows the variation of the total number density $n$ of all the hadrons in the HG with respect to baryon chemical potential $\mu_B$ at a constant temperature $T=150$ MeV. Total number density as calculated by our model does not differ much from the results of the Cleymans and Suhonen model for the initial values of baryon chemical potential. However, the difference between two model calculations becomes noticeable beyond $\mu_B=800$ MeV. Total number density as calculated in Rischke model lies far below our curve as well as Cleymans-Suhonen model results. This shows that the correction in the inconsistent model calculations arising due to the factor $\partial R/\partial\lambda$ in Eq. (37) is small. In this regard we would like to point out that we have noticed a negligible change in our results, if we cut the Neumann iterative series after $\Omega^3\,\hat R$ term in Eq. (41). This means that higher terms in Eq. (41) do not yield any appreciable difference.

     Figure 2 represents the variation of total number density with respect to temperature $T$ at constant net baryon density $n_B$. We have compared our results with those calculated by Sasaki [31] using an event generator URASiMA which is an ultra-relativistic AA collision based simulation involving the multiple scattering algorithm. He has calculated thermodynamic properties of HG by using a microscopic model (Molecular-Dynamical Simulations) that includes realistic interactions among hadrons through multiple scattering among different baryons and mesons. Total number density predicted by our model shows very close agreements with the results obtained by Sasaki. At higher temperature (i.e., $T>150$ MeV) our results and results of Sasaki differ as our model predictions lie slightly below the Sasaki results. It is also obvious from the figure that total number density increases in both the models with increase in the net baryon density at fixed temperature. However, our results indicate a rapid increase in $n$ beyond a temperature $T\approx 185$ MeV. The dependence on $n_B$ also decreases fast and the curves come closer to each other.

    Figure 3 depicts the variation of the total hadronic pressure with respect to the baryon chemical potential $\mu_B$ at fixed temperature $T=150$ MeV in different models. Hadronic pressure increases with baryon chemical potential in all the models but the prediction of our model again lies close to the thermodynamically inconsistent model of Cleymans and Suhonen. In these models, hadronic pressure shows a saturation around $\mu_B=800$ MeV. Rischke model calculation does not show any such saturation but shows monotonically increasing behaviour with respect to $\mu_B$ at fixed $T$. For comparison, we have also shown the hadronic pressure resulting from the ideal hadron gas model without any type of excluded volume corrections. The increase of the hadronic pressure in this case is more rapid. If we take the QGP equation of state as follows:
\begin{equation}
P_{QGP}=\frac{37}{90}\,\pi^2\,T^4+\frac{\mu_B^2\,T^2}{9}+\frac{\mu_B^4}{162\,\pi^2}-B,
\end{equation}
where we have considered u, d massless quarks and gluons in the EOS and $B^{1/4}=206$ MeV, then we find that the QGP pressure curve cuts the ideal hadronic pressure curve at two points i.e., when $\mu_B=200$ MeV and $\mu_B=625$ MeV, respectively. However, QGP curve cuts all other curves only at one point and hence only the phase transition from the hadron gas to QGP occurs in the excluded volume models. This illustrates the importance of excluded volume correction incorporated in the EOS of hadron gas because we do not get the anomalous reversal of phase transition from QGP to hadron gas in all such type of models.

    Figure 4 represents the variation of total hadronic pressure with respect to temperature $T$ at a fixed value of net baryon density $n_B$. Results of hadronic pressure calculated by Sasaki using event generator have also been shown by solid points. Hadronic pressure shows a very slow increase as temperature increases upto $T=170$ MeV. After this temperature pressure increases rapidly. Hadronic pressure also becomes independent of the net baryon density beyond the temperature $T>185$ MeV and this feature again matches closely with the results obtained by Sasaki.

     In Figure 5 we have shown the variation of the energy density with temperature at a constant net baryon density as calculated in our model as well as in the event generator model of Sasaki. The energy density varies slowly with temperature at a fixed net baryon density for initial values of $T$ but starts showing a sharp increase as $T$ increases and $T>180$ MeV. This trend is common in both the calculations. Our model predicts slightly lower values of the energy density as compared to results obtained in Sasaki model at all the temperatures. Also the energy density shows an appreciable dependence on the values of net baryon density at lower values of temperature. But at higher temperatures (i.e., $T>180$ MeV), energy density becomes almost independent of the net baryon density.

     Figure 6 depicts the variation of entropy per baryon $s/n_B$ with respect to the temperature at a constant $n_B$. Usually $s/n_B$ for any thermodynamic system is a constant quantity since it remains unaffected during the final stages of the evolution of the fireball. Thus we find that entropy per baryon is a measurable quantity and describes the properties of the fireball in a significant manner. Our model shows good agreement with the results obtained in the model of Sasaki. Some difference in the results is seen at lower temperatures where our model shows slightly lower values of $s/n_B$ as compared to the calculation of Sasaki. Moreover, the ratio $s/n_B$ shows a distinct dependence on   at higher values of the temperatures particularly above $T=160$ MeV. Thus the frequently used observation that $s/n_B$ is insensitive to $n_B$ is contradicted by our model and Sasaki model. It is well known fact that $s/n_B$, as predicted by ideal hadron gas model, is almost independent of $n_B$ even above $T=m_{\pi}$. Therefore, our results as well as results of Sasaki clearly indicate that the ideal hadron gas model provides a poor description for $s/n_B$ above $T=m_{\pi}$. Thus one should be more careful while using ideal hadron gas model in the interpretation of the results of ultra-relativistic heavy-ion experiments.

    In Figure 7 we have plotted the hadronic pressure versus energy density at a fixed value of entropy per particle $s/n$ where $n$ is the total number density of the hadrons. Our model predicts a linear variation of pressure with energy density. Slope of the curve gives the square of the velocity of sound in the medium and it increases with the increase in the values of $s/n$. Thus it is an important result of our model because we find that $v_s^2<1$ in the dense and hot HG. So we notice that causality cannot be violated in our excluded volume approach [22].

    In Figure 8 we have shown the variation of net baryon density with respect to the baryon chemical potential $\mu_B$ at constant temperature $T=150$ MeV.  Results of the present model calculation differ much from the Cleymans and Suhonen model if the chemical potential increases beyond $800$ MeV. Net baryon density as predicted by Rischke model lies below the curves given by our present model as well as the inconsistent model.

    Figure 9 shows the fraction of occupied volume $R$ versus baryon chemical potential $\mu_B$ at fixed temperature $T=150$ MeV in our present model, Cleymans and Suhonen model and Rischke model. The curve for $R$ in our present model lies just above the curve given by the inconsistent model. But the Rischke model calculation shows very low values of $R$ as compared to our present model.

\section{SUMMARY AND CONCLUSIONS}

   We have constructed a new thermodynamically consistent equation of state (EOS) for a hot and dense HG by incorporating the finite-size of the baryons. Here we have obtained a simple form of EOS where the excluded volume effect has been incorporated in the partition function by suitably defining the volume integral. Our present model can be suitably used even at extreme values of $T$ and $\mu_B$ which was not possible in the earlier version of our model. Our model resembles the thermodynamically inconsistent Cleymans-Suhonen model but contains extra terms as demanded by the thermodynamic consistency. We have compared the predictions of our model with those of inconsistent as well as Rischke model. We have also compared our model predictions with the predictions of microscopic model used by Sasaki. We find that our model predictions mostly show very close agreement with the Sasaki results although the two approaches are completely different in nature.  Some quantitative difference between our model and Sasaki model may be due to some extraordinary assumptions made in Sasaki model e.g., anti-baryons and strange particles are not taken into account in this model.

    In conclusion, although our results do not differ much from those of the Cleymans-Suhonen model results, yet it gives the thermodynamically consistent description of all the thermodynamic quantities like number density, pressure and energy density etc. and these are valid even for extreme values of temperatures and baryon chemical potentials. In addition it is easier to calculate in our model as compared to other thermodynamically consistent models since it does not involve any transcendental equation. We should stress here that we have given a phenomenological approach to incorporate the finite size of the baryons for hot and dense HG as 'excluded volume effect'. However, phenomenology cannot be substituted for a formal theory. Some calculations developed in the mean-field models or other kind of the theory for including excluded volume effects have appeared recently [22-24]. However, these approaches again suffer from many limitations e.g., realistic calculations involving many resonances cannot be done in these models. There are also many parameters in these theories which we cannot avoid in the calculations. We are thus confident that the EOS developed here will provide a suitable description of the experimental data obtained from lowest the lowest SIS energies to the highest RHIC energies. We intend to give the predictions of our model regarding the particle multiplicities and particle ratios and their comparison with the available experimental data in a subsequent publication.

\section*{ACKNOWLEDGMENTS}

   One of us (M. Mishra) is grateful to Council of Scientific and Industrial Research (CSIR), New Delhi for the award of Senior Research Fellowship (SRF). We are very grateful to Prof. V. J. Menon for many stimulating discussions and helpful suggestions.

\end{document}